\documentstyle[12pt]{article}
\catcode`\@=11
\long\def\@makefntext#1{ 
\protect\noindent \hbox to 3.2pt {\hskip-.9pt
$^{{\ninerm\@thefnmark}}$\hfil}#1\hfill} 

\def\thefootnote{\fnsymbol{footnote}}
 \def\@makefnmark{\hbox to 0pt{$^{\@thefnmark}$\hss}}  

\def\ps@myheadings{\let\@mkboth\@gobbletwo
\def\@oddhead{\hbox{} 
\rightmark\hfil\ninerm\thepage}
\def\@oddfoot{}\def\@evenhead{\ninerm\thepage\hfil 
\leftmark\hbox{}}\def\@evenfoot{}
\def\sectionmark##1{}\def\subsectionmark##1{}}
\evensidemargin
 0.30truein
\begin{document}

\newcommand{\symbolfootnote}{\renewcommand{\thefootnote}
        {\fnsymbol{footnote}}}
\renewcommand{\thefootnote}{\fnsymbol{footnote}}
\newcommand{\alphfootnote}
        {\setcounter{footnote}{0}
         \renewcommand{\thefootnote}{\sevenrm\alph{footnote}}}

\newcounter{sectionc}\newcounter{subsectionc}\newcounter{subsubsectionc}
\renewcommand{\section}[1] {\vspace{0.6cm}\addtocounter{sectionc}{1}
\setcounter{subsectionc}{0}\setcounter{subsubsectionc}{0}\noindent
        {\bf\thesectionc. #1}\par\vspace{0.4cm}}
\renewcommand{\subsection}[1] {\vspace{0.6cm}\addtocounter{subsectionc}{1}
        \setcounter{subsubsectionc}{0}\noindent
        {\it\thesectionc.\thesubsectionc. #1}\par\vspace{0.4cm}}
\renewcommand{\subsubsection}[1]
{\vspace{0.6cm}\addtocounter{subsubsectionc}{1}
        \noindent {\rm\thesectionc.\thesubsectionc.\thesubsubsectionc.
        #1}\par\vspace{0.4cm}}
\newcommand{\nonumsection}[1] {\vspace{0.6cm}\noindent{\bf #1}
        \par\vspace{0.4cm}}

\newcounter{appendixc}
\newcounter{subappendixc}[appendixc]
\newcounter{subsubappendixc}[subappendixc]
\renewcommand{\thesubappendixc}{\Alph{appendixc}.\arabic{subappendixc}}
\renewcommand{\thesubsubappendixc}
        {\Alph{appendixc}.\arabic{subappendixc}.\arabic{subsubappendixc}}

\renewcommand{\appendix}[1] {\vspace{0.6cm}
        \refstepcounter{appendixc}
        \setcounter{figure}{0}
        \setcounter{table}{0}
        \setcounter{equation}{0}
        \renewcommand{\thefigure}{\Alph{appendixc}.\arabic{figure}}
        \renewcommand{\thetable}{\Alph{appendixc}.\arabic{table}}
        \renewcommand{\theappendixc}{\Alph{appendixc}}
        \renewcommand{\theequation}{\Alph{appendixc}.\arabic{equation}}
        \noindent{\bf Appendix \theappendixc #1}\par\vspace{0.4cm}}
\newcommand{\subappendix}[1] {\vspace{0.6cm}
        \refstepcounter{subappendixc}
        \noindent{\bf Appendix \thesubappendixc. #1}\par\vspace{0.4cm}}
\newcommand{\subsubappendix}[1] {\vspace{0.6cm}
        \refstepcounter{subsubappendixc}
        \noindent{\it Appendix \thesubsubappendixc. #1}
        \par\vspace{0.4cm}}

\def\abstracts#1{{
        \centering{\begin{minipage}{30pc}\tenrm\baselineskip=12pt\noindent
        \centerline{\tenrm ABSTRACT}\vspace{0.3cm}
        \parindent=0pt #1
        \end{minipage} }\par}}

\newcommand{\bibit}{\it}
\newcommand{\bibbf}{\bf}
\renewenvironment{thebibliography}[1]
        {\begin{list}{\arabic{enumi}.}
        {\usecounter{enumi}\setlength{\parsep}{0pt}
\setlength{\leftmargin 1.25cm}{\rightmargin 0pt}
         \setlength{\itemsep}{0pt} \settowidth
        {\labelwidth}{#1.}\sloppy}}{\end{list}}

\topsep=0in\parsep=0in\itemsep=0in
\parindent=1.5pc

\newcounter{itemlistc}
\newcounter{romanlistc}
\newcounter{alphlistc}
\newcounter{arabiclistc}
\newenvironment{itemlist}
        {\setcounter{itemlistc}{0}
         \begin{list}{$\bullet$}
        {\usecounter{itemlistc}
         \setlength{\parsep}{0pt}
         \setlength{\itemsep}{0pt}}}{\end{list}}

\newenvironment{romanlist}
        {\setcounter{romanlistc}{0}
         \begin{list}{$($\roman{romanlistc}$)$}
        {\usecounter{romanlistc}
         \setlength{\parsep}{0pt}
         \setlength{\itemsep}{0pt}}}{\end{list}}

\newenvironment{alphlist}
        {\setcounter{alphlistc}{0}
         \begin{list}{$($\alph{alphlistc}$)$}
        {\usecounter{alphlistc}
         \setlength{\parsep}{0pt}
         \setlength{\itemsep}{0pt}}}{\end{list}}

\newenvironment{arabiclist}
        {\setcounter{arabiclistc}{0}
         \begin{list}{\arabic{arabiclistc}}
        {\usecounter{arabiclistc}
         \setlength{\parsep}{0pt}
         \setlength{\itemsep}{0pt}}}{\end{list}}

\newcommand{\fcaption}[1]{
        \refstepcounter{figure}
        \setbox\@tempboxa = \hbox{\tenrm Fig.~\thefigure. #1}
        \ifdim \wd\@tempboxa > 6in
           {\begin{center}
        \parbox{6in}{\tenrm\baselineskip=12pt Fig.~\thefigure. #1 }
            \end{center}}
        \else
             {\begin{center}
             {\tenrm Fig.~\thefigure. #1}
              \end{center}}
        \fi}

\newcommand{\tcaption}[1]{
        \refstepcounter{table}
        \setbox\@tempboxa = \hbox{\tenrm Table~\thetable. #1}
        \ifdim \wd\@tempboxa > 6in
           {\begin{center}
        \parbox{6in}{\tenrm\baselineskip=12pt Table~\thetable. #1 }
            \end{center}}
        \else
             {\begin{center}
             {\tenrm Table~\thetable. #1}
              \end{center}}
        \fi}

\def\@citex[#1]#2{\if@filesw\immediate\write\@auxout
        {\string\citation{#2}}\fi
\def\@citea{}\@cite{\@for\@citeb:=#2\do
        {\@citea\def\@citea{,}\@ifundefined
        {b@\@citeb}{{\bf ?}\@warning
        {Citation `\@citeb' on page \thepage \space undefined}}
        {\csname b@\@citeb\endcsname}}}{#1}}

\newif\if@cghi
\def\cite{\@cghitrue\@ifnextchar [{\@tempswatrue
        \@citex}{\@tempswafalse\@citex[]}}
\def\citelow{\@cghifalse\@ifnextchar [{\@tempswatrue
        \@citex}{\@tempswafalse\@citex[]}}
\def\@cite#1#2{{$\null^{#1}$\if@tempswa\typeout
        {IJCGA warning: optional citation argument
        ignored: `#2'} \fi}}
\newcommand{\citeup}{\cite}

\def\fnm#1{$^{\mbox{\scriptsize #1}}$}
\def\fnt#1#2{\footnotetext{\kern-.3em
        {$^{\mbox{\sevenrm #1}}$}{#2}}}

\font\twelvebf=cmbx10 scaled\magstep 1
\font\twelverm=cmr10 scaled\magstep 1
\font\twelveit=cmti10 scaled\magstep 1
\font\elevenbfit=cmbxti10 scaled\magstephalf
\font\elevenbf=cmbx10 scaled\magstephalf
\font\elevenrm=cmr10 scaled\magstephalf
\font\elevenit=cmti10 scaled\magstephalf
\font\bfit=cmbxti10
\font\tenbf=cmbx10
\font\tenrm=cmr10
\font\tenit=cmti10
\font\ninebf=cmbx9
\font\ninerm=cmr9
\font\nineit=cmti9
\font\eightbf=cmbx8
\font\eightrm=cmr8
\font\eightit=cmti8

\newskip\humongous \humongous=0pt plus 1000pt minus 1000pt
\def\caja{\mathsurround=0pt}
\def\eqalign#1{\,\vcenter{\openup1\jot \caja
	\ialign{\strut \hfil$\displaystyle{##}$&$
	\displaystyle{{}##}$\hfil\crcr#1\crcr}}\,}
\newif\ifdtup
\def\panorama{\global\dtuptrue \openup1\jot \caja
	\everycr{\noalign{\ifdtup \global\dtupfalse
	\vskip-\lineskiplimit \vskip\normallineskiplimit
	\else \penalty\interdisplaylinepenalty \fi}}}
\def\eqalignno#1{\panorama \tabskip=\humongous
	\halign to\displaywidth{\hfil$\displaystyle{##}$
	\tabskip=0pt&$\displaystyle{{}##}$\hfil
	\tabskip=\humongous&\llap{$##$}\tabskip=0pt
	\crcr#1\crcr}}
\def\oldrefledge{\hangindent3\parindent}
\def\oldreffmt#1{\rlap{[#1]} \hbox to 2\parindent{}}
\def\oldref#1{\par\noindent\oldrefledge \oldreffmt{#1}
	\ignorespaces}
\def\figledge{\hangindent=1.25in}
\def\figfmt#1{\rlap{Figure {#1}} \hbox to 1in{}}
\def\fig#1{\par\noindent\figledge \figfmt{#1}
	\ignorespaces}
%
\def\ie{\hbox{\it i.e.}{}}	\def\etc{\hbox{\it etc.}{}}
\def\eg{\hbox{\it e.g.}{}}	\def\cf{\hbox{\it cf.}{}}
\def\etal{\hbox{\it et al.}}
\def\dash{\hbox{---}}
\def\tr{\mathop{\rm tr}}
\def\Tr{\mathop{\rm Tr}}
\def\Im{\mathop{\rm Im}}
\def\Re{\mathop{\rm Re}}
\def\bR{\mathop{\bf R}{}}
\def\bC{\mathop{\bf C}{}}
\def\partder#1#2{{\partial #1\over\partial #2}}
\def\secder#1#2#3{{\partial^2 #1\over\partial #2 \partial #3}}
\def\bra#1{\left\langle #1\right|}
\def\ket#1{\left| #1\right\rangle}
\def\VEV#1{\left\langle #1\right\rangle}
\def\gdot#1{\rlap{$#1$}/}
\def\abs#1{\left| #1\right|}
\def\pr#1{#1^\prime}
\def\ltap{\raisebox{-.4ex}{\rlap{$\sim$}} \raisebox{.4ex}{$<$}}
\def\gtap{\raisebox{-.4ex}{\rlap{$\sim$}} \raisebox{.4ex}{$>$}}
\def\contract{\makebox[1.2em][c]{
	\mbox{\rule{.6em}{.01truein}\rule{.01truein}{.6em}}}}
\def\slash#1{#1\!\!\!/\!\,\,}
\def\beq{\begin{equation}}
\def\eeq{\end{equation}}
\def\bea{\begin{eqnarray}}
\def\eea{\end{eqnarray}}
\def\half{\frac{1}{2}}
\def\aeq{\eeq}
\def\bq{\begin{quote}}
\def\eq{\end{quote}}
\def\pr{{\sl Phys. Rev.~}}
\def\np{{\sl Nucl. Phys.~}}
\def\pl{{\sl Phys. Letters~}}
\def\prl{{\sl Phys. Rev. Letters~}}
\def \Msol {M_\odot}
\def\GeV{\,{\rm GeV}}
\def\eV {\,{\rm  eV}}
\def\Mpc{\,{\rm Mpc}}
\def\pc{\,{\rm pc}}
\def\half{\frac{1}{2}}
\def \lta {\mathrel{\vcenter
     {\hbox{$<$}\nointerlineskip\hbox{$\sim$}}}}
\def \gta {\mathrel{\vcenter
     {\hbox{$>$}\nointerlineskip\hbox{$\sim$}}}}
\def \endpage {\vfill \eject}
\def \endline {\hfill \break}
\def \etal {{\it et al.}\ }
\relax
\input epsf
\begin{flushright}
Fermilab-Pub-93/256-T\\
September 21, 1993
\end{flushright}
\vskip 1.0in
\centerline{\bf B--PHYSICS IN HADRON COLLIDERS$^\dagger$}
\vspace{1.0cm}
\centerline{\rm CHRISTOPHER T. HILL}
\baselineskip=13pt
\centerline{\it Fermi National Accelerator Laboratory}
\baselineskip=12pt
\centerline{\it  P.O. Box 500, Batavia, Illinois, 60510}
\vspace{1.0cm}
\begin{quote}
{\twelverm The possibility of exploring the systematics of the
spectroscopy, strong dynamics, and the weak and rare
decay modes of b--quark systems at hadron colliders
such as Fermilab, LHC and SSC, is discussed.  A copious yield of $10^{10}$
detected $B$--mesons is readily accessible in a dedicated
Fermilab program, and implies a vast array of accessible decay
modes, including second order weak processes and $CP$--violation,
which will be  unavailable elsewhere until the commissioning of LHC or SSC.
Kinematic and flavor tagging, utilizing the ``daughter pions'' from
resonances, is expected to play a major role in semileptonic weak decay
studies  and the search for $CP$--violation.
}
\end{quote}
\vspace{4.0cm}
\vfil

\noindent
{$\dagger$ Plenary talk, {\em Workshop on $B$ Physics at Hadron
Accelerators}, Snowmass, Colorado, June 25, 1993; Invited Lecture, {\em TASI},
Boulder, Colorado, June 18, 1993.}

\newpage
\twelverm   
\baselineskip=14pt
\section{Introduction}
\vspace*{-0.7cm}
\subsection{Generalities}
\vspace*{-0.35cm}
The $b$--quark offers a window on the  standard
model that is open to experimentalists at hadron colliders,
where the largest yields of $b$--quarks occur.
With existing facilities, such as CDF, it
should be possible to achieve $\sim 10^9$ observable $B$--decays
within the next few years. This entails evolution of the high
resolution vertex detectors, e.g.,
CDF's SVX, including full $r$-$\theta$-$z$ information, and
especially generalized triggers,
such as single lepton displaced vertices for semileptonic weak decay
studies.$^{1,2}$
With a modest yet dedicated program,
perhaps involving a new detector,
$>10^{10}$ observed $B$'s should be achievable at Tevatron
to Main Injector luminosities within this decade.
Such a program is essential to break the ground for future
hadron--based $B$-physics programs at LHC and SSC. An ultimate
hadron collider based program at Fermilab, LHC and SSC can
look forward to recording the decays of $>10^{12}$
produced $B$'s.

The present discussion is intended to be primarily a
prospectus for such a program.  We will, however, indulge in
some speculations about tagging of flavors and the all--important
kinematic reconstruction needed to do semileptonic weak studies.
This reflects recent interest that has arisen in the possibility of
``daughter pion'' tagging, i.e., using the pions from the
decays of parent resonances to tag the flavor.$^3$

The major advantages of the hadron based $B$--physics environment are
the relatively large cross--section for $b$--quark production and the
the ``broad--band'' nature of the beam.  $b$--quark pairs are produced
by  (predominantly)
gluon fusion$^4$ and arbitrarily massive states are available. Thus,
all of the spectroscopy, including $B_c\sim {\overline{b}}c$ and
the resonances, $B^{**}$ etc., are produced in hadronic collisions.
This sharply contrasts the situation in $e^+e^-$ machines that
make use of the $\Upsilon(4S)$ and $\Upsilon(5S)$ resonances in which
only the low--lying ${\overline{b}}(u,d)$ combinations can be produced.
Moreover, in $e^+e^-$ machines that operate in the continuum or on the
$Z$--peak the cross--section for $b$ production is many orders of magnitude
below that in the hadronic environment.

On the other hand these advantages imply major challenges as well.$^{1,2}$
The copious production at hadron machines implies that a substantial
parsing of data must occur quickly on--line, i.e., a trigger that can keep
interesting candidate events must be provided. To date in hadronic colliders
the semileptonic decay modes have been largely discarded in favor
of the much easier $\psi$ modes.  A trigger capable of recovering the
semileptonic decays is possible, and
demonstrating its feasibility is of high priority for a number of reasons
(conventional flavor tagging requires it).

Another issue is the extent to which decays involving missing mass,
such as the semileptonic decays involving neutrinos, can be fully
reconstructed.  In $e^+e^-$ machines that
make use of the $\Upsilon(4S)$ the $B$--mesons are produced with
a known energy, the beam energy. In combination with the visible decay
momentum, this
completely determines the decay kinematics, e.g., the $Q^2$ of
the lepton pair is determined even though the neutrino is never seen.
In a hadronic mode we observe a $B$--meson flight direction and the
visible momentum of the decay products, but this yields a two--fold ambiguity
in the $B$ energy.
Thus, to make maximal use of a semileptonic decay sample it is imperative
that efficient techniques evolve for resolving this ambiguity!

One technique
would ``bludgeon'' the semileptonic processes with high statistics by
insisting on keeping only those special kinematic configurations for
which the ambiguity disappears$^5$.  While inefficient, this
technique is guaranteed to work.
However, we will suggest another approach presently that is
speculative, but may ultimately prove to be an efficient way of fully
reconstructing
$B$ processes with relatively high efficiency.  It makes use of the
fact that $B$--mesons will often be produced as decay fragments of
a resonance as in $B^{**}\rightarrow B + \pi$.  The $\pi$ meson here we will
call a ``daughter pion,'' and it has previously
been suggested as a flavor tagging mechanisim
for neutral $B$--mesons.$^2$ The observation of daughter $\pi$--mesons from
resonances
is established by ARGUS, E-691
and CLEO, and E-687.$^6$
However, we suggest here that
it can potentially be used to resolve the two--fold kinematic ambiguity in
the $B$--meson $4$--momentum. We describe
this approach in Section 2.4 below. It may prove workable  in some form as our
understanding of $B$ production evolves.

The physics goals of a $>10^{10}$ $B$--meson program are very rich and diverse.
Heavy quark physics allows us to map out the CKM matrix of the standard
model through the detailed studies of inclusive
and exclusive decay modes. It
will allow us to test the standard model beyond
the leading order in  radiative corrections, and
through rare decay modes and mixing
phenomena  which
are sensitive to $m_{top}$ and $V_{tq}$, etc.
This will lead ultimately to  experimental tests of the
CKM theory of $CP$--violation, which is expected to
manifest itself in many interesting new channels in the $b$--system.
 High statistics studies of
the $b$--system will furthermore enable
searches for exotic physics, signals of which might be
 expected to emerge in heavy quark processes.

We begin first with a brief overview  of the physics considerations
that are relevant to doing heavy quark physics in the hadronic
collider environment.

\vglue 0.3cm
\leftline{\twelveit 1.2. Prima Facie Considerations of Hadronic $B$'s}
\vglue 1pt
$B$--physics at hadron colliders is often casually dismissed
out--of--hand, preference given to
$e^+e^-$ production, because the hadronic environment is ``too noisy.''
It is important to realize that the ``noise,''
i.e., the background of high multiplicity, mostly low
$p_T$ pions in a hadronic collision,
is largely spread out over a large range of rapidity.
The low--mass particle production follows an approximately
constant distribution in the pseudo--rapidity:
\beq
\eta = -\ln[\tan(\theta/2)] \approx \tanh^{-1}(p_z/E)
\eeq
Typically at Tevatron energies the number of pions per unit
rapidity is given by:
\beq
\frac{d N_\pi}{d\eta} \approx 3.0\;\;\;\makebox{charged};
\qquad \approx 1.5\;\;\;\makebox{neutral}
\eeq
Thus, in a rapidity range of $|\eta |<1$ we expect of order $\overline{n}
=6$ charged
pions, and $6$ $\pi^0$ gamma's emanating from the beam collision spot.

\vskip 0.2in
\begin{center}
\begin{quote}
{\tenrm Table I: Indicated yields of usable $B$--mesons running
for a 3 year, $30\%$ duty cycle, period for: (a) Tevatron at
present attainable ${\cal{L}} = 10^{31}$  cm$^{-2}$ sec$^{-1}$
(b) Main Injector assuming ${\cal{L}} = 10^{32}$  cm$^{-2}$ sec$^{-1}$
(twice the design goal; multiply by $10$ if
the rapidity range is $|\eta|\leq 3$ and
$p_t>5$ GeV). (c) ABF -- Asymmetric B-factory proposal at
${\cal{L}} = 10^{34}$  cm$^{-2}$ sec$^{-1}$ operating on the
$\Upsilon(4S)$ (d) LEP at $Z^0$--pole with
${\cal{L}} = 2\times 10^{31}$ cm$^{-2}$ sec$^{-1}$ (see M. Artuso in
ref.[2]).}
\end{quote}
\vspace{.2in}
\begin{tabular}{|| l | c | c | c | c ||}
\hline
Mode  & Tevatron$^{(a)}$  & Main Injector$^{(b)}$  &  ABF$^{(c)}$ & LEP
II$^{(d)}$ \\ \hline \hline
$B_{u,d}$ & $ 6\times 10^{9}$ & $6\times 10^{10}$ & $3\times 10^8$ & $4\times
10^{6} $ \\ \hline
$B_{s}$ & $1.6 \times 10^{9}$ & $1.6\times 10^{10}$  & none & $ 8\times 10^{5}
$  \\ \hline
$B_{c}$ & $10^{7}$ & $10^{8}$ &  none  &  $ 4\times 10^{3}$ \\ \hline
$\Lambda_b$ & $10^{9}$ & $10^{10}$  & none  &  $4\times 10^{5} $  \\ \hline
\end{tabular}
\end{center}
\vskip 0.2in

On the other hand, the finite and relatively large mass of the $b$--quark
leads to a longitudinal momentum distribution that is centered on
$\eta = 0$, and is fairly broad depending upon
the {\em cm} energy scale and the $p_t$ cut
(see, e.g., Alan Sill in ref.[1]). In rapidity,
the range of significant $b$--quark production
with high $p_t$ is for the Tevatron $\sim \pm 3$; for the
LHC $\sim \pm 4.5$; and for the SSC $\sim \pm 7$.
Moreover, the transverse momentum
distribution, $p_t$, of heavy quarks is set
by the mass scale of the quark (generally, it requires a
parton subprocess of larger $\hat{s}$ to make a heavier
quark, hence larger values of $p_t$ become relatively more
probable).

 Moreover, $b$--hadrons  have a fortuitously long life--time, and
they therefore drift a resolvable distance away from the primary vertex
before they decay. With high resolution vertex
detectors it is easy to resolve
the secondary vertex and isolate the heavy hadron decay.
The typical displacement of a $b$--hadron
secondary decay vertex is $\sim 400$ microns, while
a resolution to better than $\sim 15$ microns is achieved
with the SVX.
With this secondary vertex separation there is only a very small combinatorial
background to
these displaced vertices coming from minimum bias physics.
There remains, however, a
 significant background from charmed mesons which also have
displaced secondary vertices.  These can generally be
controlled by demanding partial reconstruction of
the heavy hadron decay with mass cuts, i.e., demand that the
visible decay products have masses exceeding those of charmed
particles, typically $\gta 2.5$ GeV.

At the luminosity of $10^{31} \; cm^{-2}\;sec^{-1}$ in a $p\overline{p}$
collider, for which we assume $\sqrt{s}=1.8$ GeV, $B$-meson pairs are
produced in a rapidity range of
$|\eta|\leq 1$ and $p_t>10$ GeV,
with a total cross-section of $\sim 10\;\mu b$ or 100 Hz
(ref.[1]; M. Artuso in ref.[2]).
With the main injector, and
the experience to date at the Tevatron, an ultimate
luminosity of $10^{32}\; cm^{-2}\;sec^{-1}$ is thinkable
(the present peak Tevatron luminosity is $\sim 0.8\times 10^{31}$). Running
at  $10^{32}$ ($10^{31}$) for
a total of 3 years, with a $33\%$ duty factor
yields $\sim 3\times 10^{10}$ ($3\times 10^{9}$)
{\em usable} $B$--mesons.  If we can triple the rapidity range
to $|\eta|\leq 3$ and reduce the lower limit to $p_t>5$ GeV the
yields for useful $B$'s approach $\sim 3\times 10^{11}$ ($3\times 10^{10}$).
Of this, the yield of $B_s$ is $\sim 18\%$, $\Lambda_B$ is $\sim 10\%$
and
of $B_c$ is $ \sim 0.1 \%$. The yield of $b$--quark
containing baryons is expected to be of order  $ 10\%$,
though these are crude estimates at present, and should
actually be measured at the end of run I.

This compares with the idealized luminosity of $10^{34}
\; cm^{-2}\;sec^{-1}$ in  an $e^+ e^-$ storage ring, such as
the proposed asymmetric $B$-factory (ABF) at SLAC or CESR
(the present peak luminosity at CESR is $2.5\times 10^{32}$). The cross-section
for $B\overline{B}$ production on
the $\Upsilon(4S)$ is $\sim 1 $ nb, which yields
$B$ pairs on the $\Upsilon(4S)$ at a rate of $\sim 10 $ Hz.  The yield for
the same 3 year $30\%$ duty cycle
period is $\sim 3\times 10^8$ $B$--mesons
(note this is  the proposed ultimate
$300$ $fb^{-1}$, lifetime $\int{\cal{L}}dt$ for the asymmetric
$B$--factory). On the
$Z^0$ pole the cross--section is $\sim 7.0$ nb. Hence operating
an $e^+e^-$ collider at the
LEP  luminosity
of $2 \times 10^{31}$  on the $Z^0$ pole
for the same 3 year continuous duty cycle
period yields $\sim 3\times 10^{6} $ $b$'s.  For continuum
$e^+e^-$ machines the cross--sections are $\sim 10^{-2}$
those on the $Z^0$ pole and we will not consider them
for comparison.

We see from Table I that
various new states and
decay modes are available in the hadronic
facility that are inaccessible, or of lower statistics in
the $e^+ e^-$ environment.
Moreover, it appears that a reasonable goal for a dedicated hadron
collider based program in the pre--SSC era is to produce a total
of $>10^{10}$ usable $b$ hadrons.  In what follows we will take $10^{10}$
$B$--mesons  to
be our
standard reference normalization and  give a
preliminary consideration of what might be achieved in such a
program.

\section{Physical States and Leading Processes }

\vglue 0.3cm
\leftline{\twelveit 2.1. Resonance Spectroscopy}
\vglue 1pt

The spectrum of resonances of the $B$--mesons imitates that
of the charm system.
We see this by comparison in Fig.(1), where the known and
predicted resonances of $\ell=0$ and $\ell=1$ are indicated.
The spectroscopy is actually reflecting a remarkable
aspect of  heavy quark symmetry, i.e.,
the heavy quark spin symmetry.$^7$

Put simply, heavy quark spin can be ignored
in the dynamics, and acts effectively  like a flavor symmetry.  As a
result, states which differ only by flipping the heavy quark spin will be
 degenerate (up to $O(1/M)$).
It is convenient to describe this by classifying mesons as
$(j_1,j_2)$, where $j_1$ is the spin of the heavy quark subsystem
and $j_2$ is the spin of the remaining system.  So, for a heavy--light
meson $j_1=\half$, and the states of lowest mass will have $j_2=\half$
as well.  Thus $(\half,\half)$ describes the groundstate and this corresponds
to total $J=0$ or $J=1$.  Therefore, the goundstate consists
of a degenerate $0^-$ and $1^-$ multiplet.  We see the $D$ and $D^*$
are actually split by slightly more than
a pion mass, while the
splitting between the $B$ and $B^*$ decreases by $m_c/m_b$ in the $B$ system.
It is important to note that $j_2$ is the quantum number of the ``brown
muck;'' we cannot {\em a priori} separate the light quark and gluon degrees of
freedom under rotations in QCD, though potential models
do so (potential models refer to constituent quarks, and work remarkably
well even in light heavy--light systems).$^8$
A fancier way of stating this is to note that spin is the
classification of a state under the ``little group;''  the little group
is the subgroup of the Lorentz group which commutes with the momentum
of the state (i.e., it is just $O(3)= SU(2)$ for a massive
particle, or $O(2)=U(1)$ for a massless particle).  Remarkably, we see that
the
little group of a heavy--light meson is enlarged to $SU(2)\times SU(2)$,
since we can rotate the heavy quark independently of the brown muck.
The states for which $|j_1-j_2|$ is an integer are equivalent to
representations of $O(4)=SU(2)\times SU(2)$. Thus the groundstate
is equivalent to a {\bf 4}-plet under $O(4)$, containing the
$0^-$ and the $1^-$ mesons.

 \vskip .5in
\hspace{-0.5in}$
\begin{array}{cc}
{\epsfxsize=3.0in
\epsfysize=3.0in
\epsffile{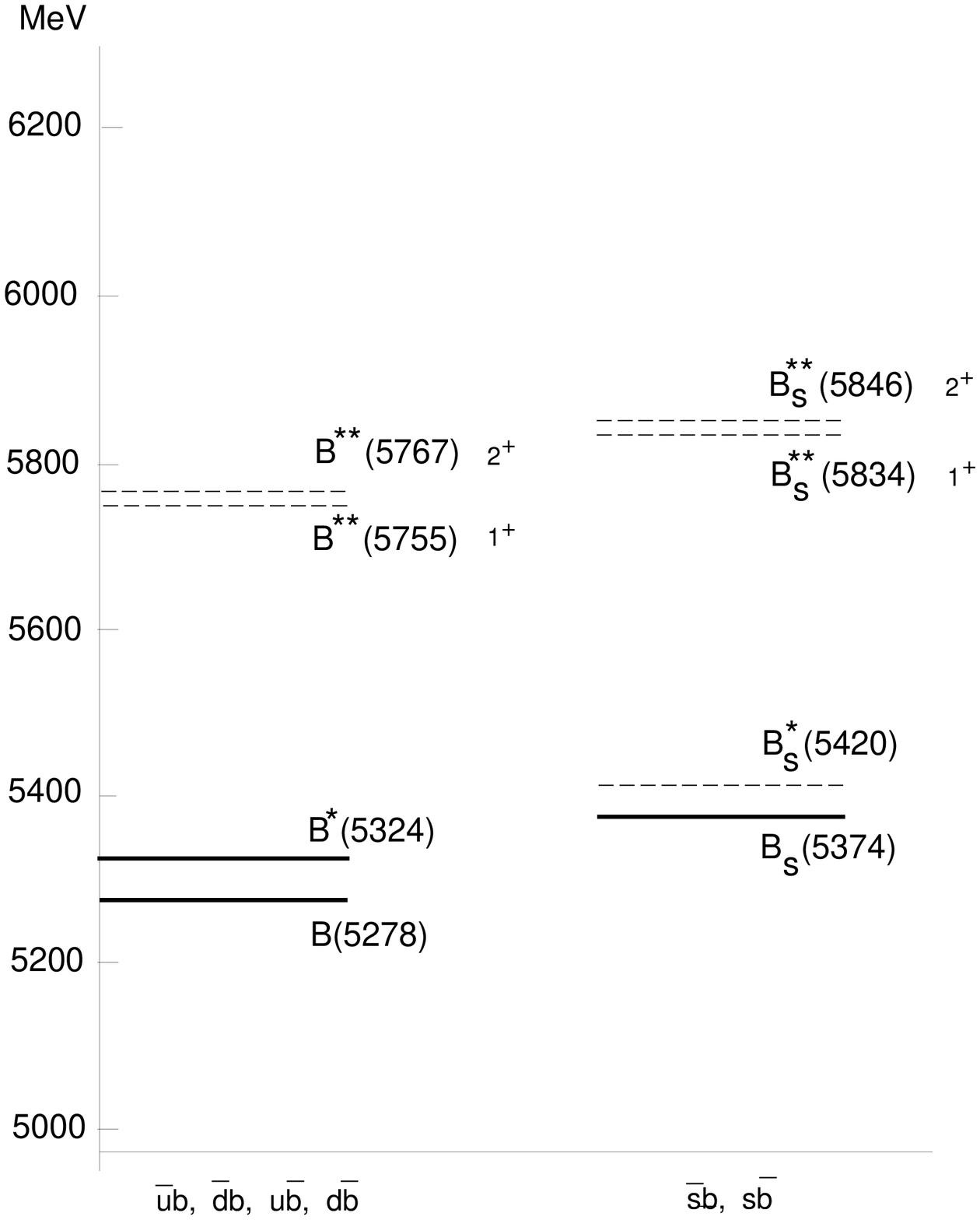}}
 &
{\epsfxsize=3.0in
\epsfysize=3.0in
\epsffile{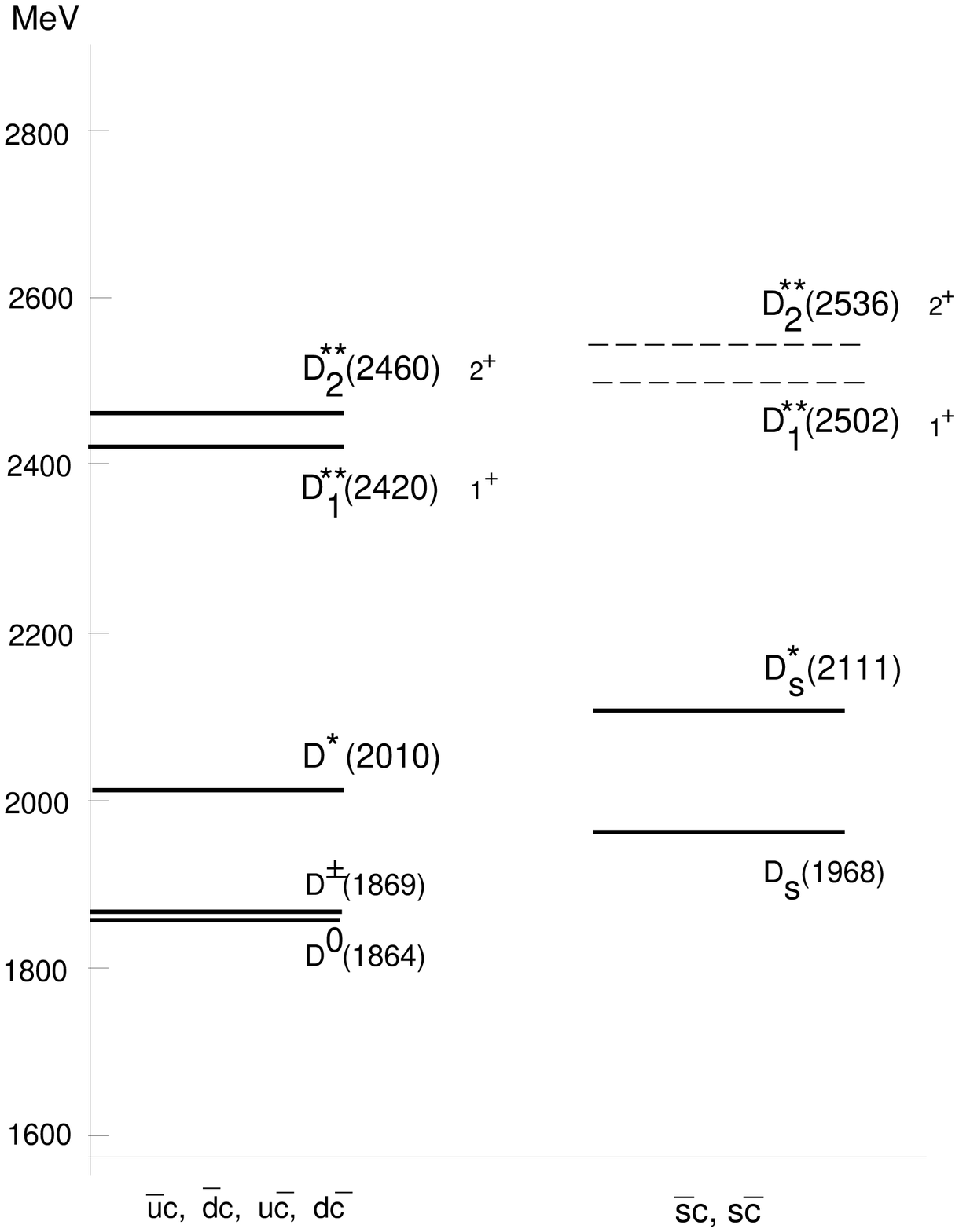}}\end{array}
$
 \vskip .1in
\begin{quote}
{\tenrm Figure 1. The low--lying spectra of $D$ and $B$
states from EHQ.$^8$ Solid lines are
established, dashed lines are predictions (we omit the broad
$(0^+,1^+)$ p-waves$^9$).}
\end{quote}
\vskip .1in

The masses and decay widths of heavy--light resonances
have been estimated
recently  by Eichten, Hill and
Quigg (EHQ)$^8$.  The
masses of these states seem to be well fit by using a Buchm\"{u}ller--Tye
potential for a static massive quark with a constituent light quark
boundstate.  Their decay widths were obtained  by  rescaling the
known strange and charm widths  with smearing.  The spectra are presented
in Fig.(1). There will generally occur a $(\half,\half) =0^+ + 1^+$
parity partner of the groundstate (a $p$--wave in the constituent quark model)
which has a very large $\sim GeV$ width and will generally be unobservable.$^9$
This state may be viewed as the ``chiral partner''
of the groundstate;$^9$ if we imagine restoring the broken chiral symmetry the
groundstate would have to linearly realize the chiral symmetry, thus becoming
doubly degenerate (thus, the left--handed iso--doublet is $0^+-0^-$, while
the right--handed iso--doublet is $0^++0^-$ when the chiral symmetry is
restored).

\vglue 0.3cm
\leftline{\twelveit 2.2. Daughter Mesons}
\vglue 1pt

The resonances can be observed by studying the $\pi$'s
and $K$'s produced in association with $B$--mesons.  Some of the
$\pi$--mesons will be decay relics from processes like:
\beq
p + \overline{p} \rightarrow X + (B^{**}\rightarrow B + (\pi,K))
\eeq
The first objective is to establish the existence and masses
of the resonant states and the fraction $f = \sigma_{B**}/\sigma_B$ by which a
$B$--meson is produced through the decay of parent resonance.
$f$ is likely to be sensitive to the decay and production
kinematics.

Experience in $e^+e^-$ (ARGUS and CLEO) and  charm
photo--production experiments$^{6}$ suggests
$f\sim 13\%$ for the fraction of $D^*$ coming
from the $D^{**}$, and $f\sim 7\%$ for the fraction of $D$ coming
from the $D^{**}$. We note that photoproduction on
a hadronic target (E-691, E-687 in ref.[6]) bears some
formal resemblance to the gluon fusion
process, and might be a good analogue process for calibrating
our understanding of detailed
production in $\overline{p}p$ collisions.  We would expect
(heavy quark symmetry) that apart from normalization the
charm production distributions can be taken over to
$B$-physics directly.
Thus, for tagging purposes an inclusive
rate of $f\sim 20\%$ of $B$'s coming from the $B^{**}\rightarrow
B^*\rightarrow B$ and  $B^{**}\rightarrow B$ chains might be
expected. The experience in photoproduction suggests that the
efficiency for finding the daughter pion is $\sim 50\%$.  We will
therefore assume an overall tagging efficiency of $\sim 10\%$ by
daughter mesons is possible.

The production tagging efficiency is probably sensitive to $p_T$ and
to angular cuts (or rapidity cuts).  The heavy quark limit ensures that the
$4$--velocity of the
produced $B^{**}$ is approximately equal to the $4$--velocity of the
$B$ i.e., zero--recoil of the $b$--system is a good approximation.
In hadronic collisions
it is probably reasonable to assume that the $B^{**}$ system at low $p_T$ is
produced in an unpolarized initial state and, thus,
the distribution of decay pions in
the process $B^{**}\rightarrow B+\pi$ is spherical in the
$B^{**}$ rest frame. The (unit normalized)
polar distribution of pions relative to the
$B$ flight direction is then obtained by boosting the
spherical distribution:
\beq
\frac{dN}{d\Omega} =
\frac{1}{4\pi}\left[
\frac{\gamma(1-\beta^2\omega^2-((\beta^4-2\beta^2)\omega^2+\beta^2)
\cos^2\theta) + 2A\beta\omega\cos\theta}{A\gamma^2(\beta^2\cos^2\theta-
1)^2}\right]
\eeq
where:
\beq
A = (1-\beta^2\omega^2 - \beta^2(1- \omega^2)\cos^2\theta)^{1/2}
\eeq
and $\omega = 1/ \sqrt{1 -m_\pi^2/(\Delta M)^2}\approx 1.04$. In
the massless pion limit, $\omega=1$ and this reduces to
${dN}/{d\Omega} =
{1}/[{4\pi\gamma^2}{
(1 - \beta\cos\theta)^2}]$
valid to order $\half O(m_\pi^2/(\Delta M)^2)
\approx 4\%$. Note that
$50\%$
of the pions will occur within a cone of opening angle
$\theta_{50\%}$ given by (for $\omega =1$):
\beq
\tan\theta_{50\%}  = \frac{1}{\gamma\beta}
= \frac{1}{\sqrt{\gamma^2 - 1}}
\eeq
For $\gamma\approx 2$ we see that $\theta_{50\%} \sim 30^o$, and
this defines a cone of small solid angle of $0.07\times 4\pi$
steradians.  The aligned daughter pions, coming from the primary vertex,
are also expected be more energetic than typical minimum
bias pions.  Thus, the conical cut on pions with
$\theta < \theta_{50\%}$ should lead to a significant gain
in signal to background for low--$p_T$ (at high $p_T$ the
$B$--meson is enveloped in a jet with higher $\pi$
multiplicity within small conical angle).   We do not consider the more general
possibility of rapidity correlations here.$^3$

\vglue 0.3cm
\leftline{\twelveit 2.3.Semileptonic Weak Decays involving $V_{cb}$}
\vglue 1pt

High statistics measurements of exclusive
semileptonic branching ratios
such as $B\rightarrow l +  \nu + (D^{**}, D^*, D)$,
etc.,  are possible at the level of $\sim
10^9$ decays.  These are important processes for establishing the overall
normalization of weak transitions in hadron colliders since the
CLEO and ARGUS experiments are significantly improving the statistics of
these processes.
The key physics goal here is to obtain the highest precision determination
of $V_{cb}$ possible.  This requires exploiting the heavy quark
symmetry result, together with QCD and $1/M$ corrections, which fixes
at special kinematic point $w= v\cdot v'\rightarrow 1$
the normalization of the Isgur--Wise function.
The normalization of $\xi(w\rightarrow 1)$ is known to a precision approaching
$3\%$.$^{10}$  Therefore, the goal of experiment should
be to approach a $3\%$ determination of $V_{cb}$.

Much effort to date has gone into the measurements of
 semileptonic weak inclusive decays and
exclusive  decays of heavy mesons.
In $e^+ e^-$ experiments such as CLEO or ARGUS, and as proposed
for the asymmetric $B$--factory, one tunes the beam energy
to produce the $\Upsilon(4S)$ resonance,
which decays to pairs of $B^+ {B}^-$ or $B^0 \overline{B}^0$
mesons that are nearly at rest in their {\em cm} system. With tagging
this can produce a  clean sample of $B$'s for the
exclusive decay modes.
The $B$--mesons can then decay
to a final state lepton either directly,
semileptonically as $B\rightarrow (l \nu) X$, or hadronically,
cascading as $B\rightarrow  X \rightarrow (l\nu)X'$.

Various models,$^{11}$
are used to fit the leptonic energy distribution
to the various component subprocesses (see discussion of
S. Stone in ref.[2]).
The error in these
results is dominated by the theoretical models
used to fit the spectra,
and is of order $\sim 15\%$.
At PEP, PETRA and the LEP experiments the  semileptonic
decays are studied at much higher energies. These results are consistent with
the $\Upsilon(4S)$ results to order $\sim 15\%$.$^2$
Alternatively,  one
can study exclusive modes
using a tagged $B$, and
determine the missing $M^2$ distribution from
the mass of the visible decay fragments of the other $B$. The
missing $M^2$ distribution will contain endpoint
peaks from contributing subprocesses,
such as $B^0\rightarrow l^-\nu(D^{*+})\rightarrow
\pi^+ (D^0)\rightarrow K^-\pi^+$. The subprocesses
are  then fit to the observed missing $M^2$ distribution.
These  methods, using different theoretical
models, have broadly consistently
yielded our first determinations of branching ratios
and again yield results to order $\sim 15\%$.$^{2,11}$

However, ultimately we want to minimize the sensitivity to
theoretical models in extracting $V_{cb}$, $V_{ub}$.  Here we
can use heavy quark symmetry
in a model independent way,$^{10}$ from the $w$
distribution.
The decay distribution in $w$ for
$B\rightarrow \ell\nu D_i$ is:
\beq
\frac{d\Gamma_{D_i}}{d w} = \frac{G_F^2}{48\pi^3}|V_{cb}|^2 m_B^2 m_{D}^2
(1+w)^2 \sqrt{w^2-1} \left( F_{D_i}(r,w)\right)
\eeq
where $r=m_{D_i}/m_B$ and $F_{D_i}(r,w)$ is a form factor.$^{10}$ In the
$m_{B,D}\rightarrow \infty$ limit $F$ is given in terms of the
Isgur--Wise function $\xi(w)$ and the known ratio $r$.
At the special ``zero recoil'' point
$\xi(1)=1 + \epsilon$ where $\epsilon $ is composed of
(a) QCD corrections computed to NLLA order $\pm 1\%$ and
(b) $1/M$ effects that are dominant $\pm 3\%$.
Hence, the strategy is to extract the functional dependence of
$F(r,w)$, or $\xi(w)$ upon $w$ and extrapolate to $w=1$ where
theoretical corrections are under control.  This implies that
the experimental statistical uncertainties must become significantly
smaller than $\sim 1\%$ and the limiting attainable precision of
$V_{cb}$ is expected to be $\sim 3\%$, modulo improvements in
the theoretical uncertainties.

Neubert$^{10}$ has carried out this analysis with the existing CLEO
and ARGUS data on the $q^2$ distributions, based
upon $\sim$ (a few $100$) events, to extract the model independent
result $|V_{cb}| = 0.042\pm 0.007$. This is  indicative of the
current statistical extrapolation errors
attained with $\sim 300 $ events, and this should improve in the
near future.
It would appear that with $10^4$ fully reconstructed
events the statistical error
in this approach will scale downward by a factor of $10$.
The key
point  here is that the theoretical modeling in the hadronic environment
is now relegated to the corrections, and not to the result
itself. The highest experimental statistics will drive the future
determinations of $V_{cb}$.

The challenge for this approach in the hadronic experiments is
the requirement
to fully reconstruct the decaying $B$--meson, particularly with
respect to kinematics.   In $e^+ e^-$ experiments the the beam energy,
together with the flight direction of the $B$, supplies sufficient kinematic
information to know
the $B$ energy unambiguously.  In the broad--band hadronic environment
we are {\em a priori} limited to knowing only the flight $3$--vector of the
$B$,
and the visible  $4$--momenta;
the unobserved neutrino momentum leads to the ambiguity.

Let us consider the semileptonic decay $B\rightarrow D + \ell^\pm
+ X$. Of course, $X$ contains the neutrino but may also
contain missing neutrals as well.
The first question is,
can we select events in which $w\rightarrow 1$ using this
information alone? If we consider events for
which {\em we hypothesize that} the missing (mass)$^2$ is $M_X^2$, then the
energy of the $B$ is determined up to a
a two--fold ambiguity.
\beq
E_B = \frac{\Delta^2 E_{vis }\pm
[\Delta^4 E^2_{vis } - (E^2_{vis} - \vec{p}_{vis}^{\; 2}\cos^2\theta)
(\Delta^4 + M_B^2\vec{p}_{vis}^{\;2}\cos^2\theta)]^{1/2} }{(E^2_{vis} -
\vec{p}_{vis}^{\;2}\cos^2\theta) }
\eeq
where $(E_{vis}$, $\vec{p}_{vis}) = p_{vis}^\mu = p_D^\mu + p_\ell^\mu$ is the
visible $4$--momentum ($M_{vis}^2=p_{vis}^2$) and  $\Delta^2 = \half(M_B^2 +
M_{vis}^2 - M_X^2)$.

\vskip 0.2in
\hspace{0.5in} {\epsfxsize=4in
\epsfysize=2in
\epsffile{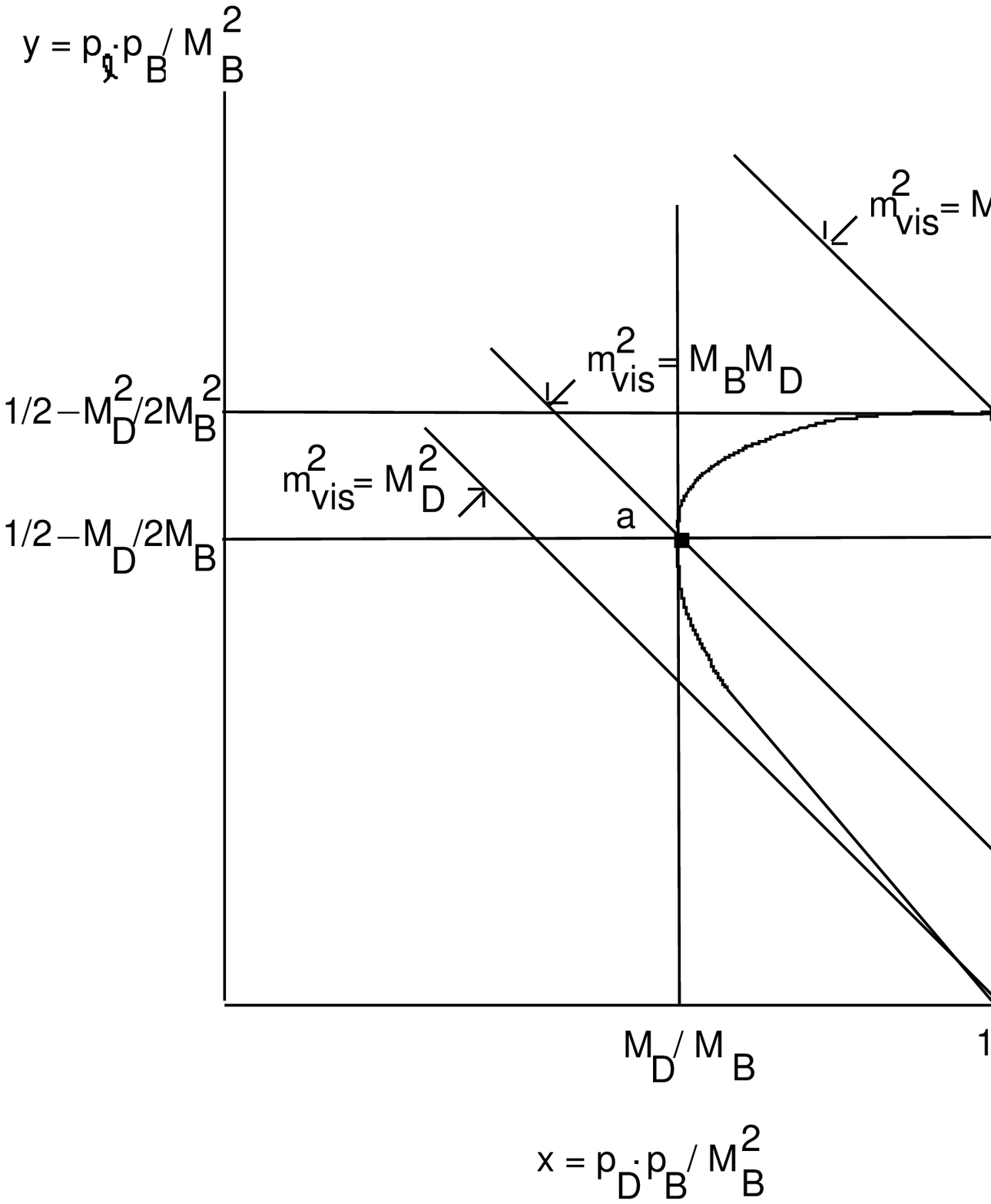}}
\begin{quote} {\tenrm Figure 2: The phase space for $B\rightarrow D + \ell +
\nu$
($M_X^2=0$) in
the variables $x\sim E_D/M_B$, $y\sim E_\ell/M_B$.  The phase space is
bounded by the points (a) ($\ell$ and $\nu$ back-to-back), (b)
($D$ and $\ell$ back-to-back), and (c) ($D$ and $\nu$ back-to-back).
The point (a) corresponds to  $w=v_D\cdot v_B = 1$.}
\end{quote}

\vskip 0.2in
\noindent
$\theta$ is the angle subtended by the flight vector of the $B$
(primary to secondary vertex vector) and $\vec{p}_{vis}$.  Let us now further
assume $M_X=0$ (no missing neutrals, etc.).
To observe $w=v_D\cdot v_B=1$ we must have in the $B$ rest--frame,
$M_B=M_D + 2E_\ell$, i.e., the massless leptons are back--to--back, whence
\beq
M_{vis}^2 = (p_D+p_\ell)^2 = M_D^2 + 2M_DE_\ell = M_D M_B
\eeq
The  condition $M_{vis}^2  = M_D M_B$, using,
\beq
0=M_X^2
= (p_D+p_\ell -p_B)^2
\qquad \makebox{implies}\qquad  x+y= \half (1+ M_{D}/M_B),
\eeq
which defines a line in the phase space
of the  decay Fig.(2) intersecting point (a).
Unfortunately this line cuts accross the
physical region (interior to (abc)) and
does not uniquely select $w=1$, while $M_{vis}^2=M_D^2$ and
$M_{vis}^2=M_M^2$ do uniquely select points (b) and (c).
Thus, for $w=1$:
\beq
E_B = \frac{ E_{vis} }{2}\left(\frac{M_B}{M_D} + 1\right)\pm \frac{
|\vec{p}_{vis}| }{2}\left(\frac{M_B}{M_D} - 1\right)
\eeq
Therefore, we see that we cannot uniquely reconstruct the
Isgur--Wise point $w=1$ from $M_{vis}^2$ alone.
To uniquely reconstruct the kinematic point $w=1$ using the information about
the $B$ decay alone
we must have (i) $M_X^2=0$ (ii) $M^2_{vis} = M_DM_B$, (iii)   and
$|\vec{p}_{vis}|=0$. Note that for $|\vec{p}_{vis}|=0$ the $B$--energy
is determined uniquely as $E_B=\Delta^2/E_{vis}$.

P. Sphicas$^5$
has examined by Monte Carlo the fraction of
(hypothetical) events for which the two--fold energy
ambiguity of the $B$--meson is less than $10\%$.
For $\sim 10^9$ decays he finds
(few)$\times 10^3$ decays in which $\delta E_B/E_B < 10\%$.
The slope of the Isgur--Wise function near
$v\cdot v'=1$ is $\xi'/\xi\sim -0.4$, thus a $10\%$ precision in
the $B$ energy yields about an additional $4\%$ uncertainty in the
normalization, or about $\sim 6\%$ overall. With $10^{11}$ $B$'s this
would approach the desired limiting resolution.

How well does this do in excluding missing neutrals?
If we allow $M_X^2=m_\pi^2 $, which occurs for a fast pion
collinear to the neutrino, then one finds that the point
(a) shifts by $\delta x_a \sim \delta y_a \sim O(m_\pi^2/M_B^2)$
(the points (b) and (c) shift by $O(m_\pi/M_B)$, which is
easier to resolve).
This is much less than the experimental momentum resolution, and
is therefore problematic. However, the typical pion contribution
is not collinear with the neutrino and $M_X^2\sim m_\pi M_B $,
whence $\delta x_a \sim \delta y_a \sim O(m_\pi/M_B)\sim 3\%$,
and is marginally resolvable.

\vglue 0.3cm
\leftline{\twelveit 2.4. Kinematic Tagging with Daughter Pions?}
\vglue 1pt

Let me indulge here in a speculative proposal.  Clearly we can
sacrifice the huge statistics available at the hadron machine to
achieve reasonable kinematic reconstruction for a (few)$\times 10^3$
events.  However, we would prefer a method which is efficient, covers
all of phase space, not just $\vec{p}_{vis}=0$,
and ideally which offers greater leverage in momentum resolution.

Perhaps we can exploit the fact that a fraction $f\sim 20\%$ of $B$--mesons
will be produced as the daughters of the  $B^{**}$ resonance, together with
the daughter pion.  Thus, let us ask if we can select
the $B$--meson energy in a typical process
$B\rightarrow D^* + \ell + \nu$,
where the two hypothetical 4--momenta of the $B$ are $p_\mu^{(1)}$,
$p_\mu^{(2)}$.
We demand that we find a pion which matches a hypothetical solution for the
$B$--meson $4$--momentum, ${p}_B$, satisfying either:
\beq
(p_\pi +  {p}_B^{(1)})^2 = M_{B**}^2\qquad
\makebox{or} \qquad
(p_\pi +  {p}_B^{(2)})^2 = (M_{B**}+ \delta M_{B**}) ^2
\eeq
where $\delta M_{B**}$ is the width of the resonance parent. Then
a difference between the hypothetical 4--momentum has a resolution
given by the width:
\beq
p_\pi^\mu({p}_B^{(1)}-{p}_B^{(2)})=  M_{B**} \delta M_{B**}=
 E_\pi (\delta_r E_B) (1 - (1 + \beta - \beta^2)\cos\theta)
\eeq
where $\delta_r E_B$ is the {\em minimum resolvable $B$--energy}.
Hence, apparently we can directly reconstruct the $B$ energy
by this method to a limiting resolution of only:
\beq
\frac{\delta_r E_B}{M_{B**}} \sim \frac{\delta M_{B**} }{E_\pi} \gta 5\%
\eeq
where we use $E_\pi \sim 1$ GeV, $\delta M_{B**}\sim 50$ MeV, typically, and
$\theta \approx 90^o$.
On the other hand, we see in eq.(11) that,  using $\vec{p}_{vis}$ the energy
ambiguity is:
\beq
\delta E_B = |\vec{p}_{vis}|\left(\frac{M_B}{M_D}-1\right)
\eeq
Note that $|\vec{p}_{vis}|$ can be quite large; as we approach
the Isgur--Wise point (a) in Fig.(2)
and, taking for example
the $B$ rest--frame, we have $|\vec{p}_{vis}|
\sim \half(M_B-M_D)$.
The value $\delta_r E_B$ is then sufficiently small
to allow a selection between the
two solutions, since:
\beq
\frac{\delta_r E_B}{\delta E_B }\sim \frac{M_{B**}\delta M_{B**}
}{E_\pi|\vec{p}_{vis}|}\left(\frac{M_B}{M_D}-1\right)^{-1} \sim 10\%
\eeq
using $|\vec{p}_{vis}|\sim \half(M_B-M_D)$. In other words,
the energy ambiguity can be $\sim 10\sigma$ of
the minimum resolvable energy
of the $B$--meson, using the daughter pion in combination.

Note that we are not then restricted to the special
kinematic configurations $|\vec{p}_{vis}|=0$; indeed, this approach
would be complimentary to $|\vec{p}_{vis}|=0$, and preferably requires that
$|\vec{p}_{vis}|$
be large.  It does rely on being able to ``cut hard'' to reduce
the background pions that fake a $B^{**}$ daughter, and it
is subject to background fakes that favor the wrong solution.
This probably favors
low $p_T$ $B$'s with less of an enveloping jet structure, and then
a $<\theta_{50\%}$ cut.
Again, this cannot resolve the missing collinear pion ambiguity, but it is
potentially able to resolve the typical missing neutral pion
ambiguity.
We have given here only a sketchy analysis
of this. It requires serious study by Monte Carlo simulation, or direct
application to the existing data of charm photoproduction experiments,
and eventually in $B$ decays where the $B$--momentum is known (all decay
products visible). With $f\sim 10\%$ we may
hope to be able to select between kinematic options with efficiencies
of order $1\%$, allowing $\sim 10^7$ fully reconstructed semileptonic
weak decays.

\vglue 0.3cm
\leftline{\twelveit 2.5. Semileptonic Weak Decays involving $V_{ub}$}
\vglue 1pt

High statistics measurements of exclusive
semileptonic branching ratios
such as $B\rightarrow \ell +  \nu + (\overline{u},(d,s))$,
etc.,  are possible at the level of a (few)$\times
10^6$ decays.  These are important processes to establish the
general
normalization of weak transitions involving $V_{ub}$.
The statistical limitations together with theory
imply better than a $3\%$ determination of the quantity
$f_B\sqrt{B}V_{ub}$ may be possible.  The quantity $f_B\sqrt{B}$ is known
poorly to about $20\%$ precision, implying an overall
determination of $V_{ub}\pm 20\%$.

The present determinations of  $V_{ub}/V_{cb}$ are
 based upon the endpoint of
the lepton spectrum  for inclusive semileptonic decay rates (see
the S. Stone review in ref.[2]).
There have been searches for the exclusive decay mode $B\rightarrow
\rho \ell\nu $.  On the
$\Upsilon(4S)$ the measurement of $E_\ell$ near the endpoint
where the background from $b\rightarrow c \ell \nu$ and continuum
$e^+e^-$ production becomes small in principle yields
a determination of $V_{ub}/V_{cb}$, however it is subject
to limitations from the knowledge of $m_b$ and $m_c$, and
is highly model dependent. The extracted  $V_{ub}/V_{cb}$
values range from $0.11\pm 0.02$ for the ACM model to
$0.17\pm 0.03$ for the ISGW model. The statistical errors
are large.
The exclusive decay mode $B\rightarrow \rho \ell \nu$ has been
studied, with greater model dependence, lower
statistics $< 100$ events
and a larger scatter of $0.1< V_{ub}/V_{cb}< 0.3$.

\vskip 0.1in
\begin{center}
\begin{quote}
{\tenrm Table II. Branching ratios estimated by rescaling charm analogues,
assuming $|V_{bu}/V_{bc}| = 0.05$.
The yields assume $33\%$ $B^\pm$, $33\%$ $B^0$, $18\%$ $B_s$. }
\end{quote}
\vskip 0.1in
\begin{tabular}{|| l | c | c |c ||}
\hline
Mode  & Br & yield$/10^{10}$ $B$'s & comment  \\ \hline \hline
$B\rightarrow \rho l \nu \; (\rho\rightarrow \pi^+ \pi^-)$
& $5.0\times 10^{-5}$ & $1.5\times 10^5$ & $\star$  lattice \\ \hline
$B\rightarrow X_{charmless} l \nu \; $ & $2.5\times 10^{-4}$
 & $1.7\times 10^6$ & inclusive models\\ \hline
$B\rightarrow \omega l \nu \; (\omega\rightarrow \pi^+ \pi^-)$
& $1.0\times 10^{-6}$ & $3.3\times 10^3$ & \\ \hline
$B\rightarrow \phi l \nu \; (\phi\rightarrow K^+ K^-)$
& $2.7\times 10^{-7}$ & $8.3\times 10^2$ & \\ \hline
$B\rightarrow \pi l \nu $ & $3.0\times 10^{-5}$ & $ 10^5$ & $\star$ chiral
symmetry
\\ \hline
$B\rightarrow \eta l \nu \; (\eta\rightarrow \pi^+\pi^-e^+e^-) $
& $1.0\times 10^{-7}$ & $ 10^3$ & chiral symmetry
\\ \hline
$B\rightarrow D_s(\pi,\rho,\omega) $ & $ 10^{-4}$ & $ 10^6$ & Argus limit $\lta
1\%$\\ \hline
$B_s\rightarrow K l \nu $ & $ 3\times 10^{-5}$ & $6\times  10^4$ &
$\star$  yields $f_{B_s}/f_{B_{(u,d)}}$ \\ \hline
$B_s\rightarrow K^* l \nu $
& $ 5\times 10^{-5}$ & $ 10^5$ & $\propto f_{B_s}$ \\ \hline
\end{tabular}
\end{center}

\vskip 0.1in
With a reasonable extrapolation to the SVX technology,
and the  copious yield of $B$'s we can imagine rather conservative cuts
allowing the study of final states such $X=\rho$, $X=\pi$,
$X=\omega$, $X=$ many $\pi$'s, etc.
In the decay $B^-\rightarrow \rho \ell^- \nu$ and the subsequent
$\rho\rightarrow \pi^+\pi^-$ ($P=0.5$) we demand that the pions
reconstruct to the $\rho$ mass, and connect to the lepton at
the decay vertex of the $B$.  The estimated
$Br(B^-\rightarrow \rho l^- \nu) \sim (Br(B^-\rightarrow D^* l^- \nu)\sim 4\%)
\times |V_{bu}/V_{bc}|^2\times 1/2\sim
5.0\times 10^{-5 }$, thus with $10^{10} $ produced $B$'s we will have $ \sim
1.5\times 10^{5}$ events.
The problematic backgrounds are from $B\rightarrow
D \ell \nu$ and $D\rightarrow 2\pi$ or $D\rightarrow \rho \pi^0$,
with the $\pi^0$ undetected,
$B\rightarrow \rho D$ and $D\rightarrow \ell \nu $.
The $\rho$ tends to be diluted by the pion background,
which may require cutting on events in which the other $B$
is seen in a semileptonic mode ($\sim 10\%$).
The rejection of $\gamma$'s and the mass reconstruction of
the $\rho$, and a veto on more than
$2$ pions are important constraints to consider in
fishing the $\rho$ out of hadronic events.

Thus, a high statistics study of Cabibbo suppressed decay modes seems possible
with $10^{10}$ $B$--mesons, but
we are in a learning situation at present that must
evolve considerably.  This yields of order $10^5$
decays.  A form factor analysis may be possible for
the $\pi \ell \nu$ mode if daughter pion kinematic tagging is possible,
yielding $\sim 10^3$ fully reconstructed decays. One can
hope to exploit the fact that
chiral symmetry fixes the normalization of this
 matrix element at $w=1$.
It should certainly be possible to achieve $V_{ub}$ to better
than $\pm 20\%$ using models, and perhaps better precision by use of chiral
symmetry. The quantity $f_{Bs}/f_{Bu,d}$ would be probed to $\pm 1\%$
precision.

\vglue 0.3cm
\leftline{\twelveit 2.6. $B_s$ and $B_c$ }
\vglue 1pt

The $B_s = (\overline{b}s)$ has been seen at Aleph, Opal and
CDF.$^{12}$  CDF has observed $14$ fully reconstructed $\psi \phi$ events,
and reports a mass of $M_{Bs} = 5383\pm 7$ MeV.  With a yield of $10^{10}$
usable $B$'s there are expected to be produced $1.8\times 10^{9}$
$B_s + \overline{B}_s$.  This will allow survey of various decay modes,
such as $DK^*$, $D^* K$, $D_S^* D_S^*$,
$D_S^*\ell\nu$, etc.  Also, of great interest will be the study of
higher resonances producing daughter $K$--mesons in association
with the $B_s$, e.g.,
\beq
p \overline{p} \rightarrow B^{***}_u(2^-,3^-) \rightarrow K + B_s
\eeq
The prospects for the application of this to, e.g.,
flavor tagging for study of
$B_s \overline{B}_s$ mixing, is discussed below.

\begin{center}
\begin{quote}
{\tenrm Table III. (a) Yields are for detectable decays and include the
branching
fractions  $\psi\rightarrow \mu^+\mu^-\sim 7\%$ (b) includes
$(\psi\rightarrow \mu^+\mu^-)\times(D^\star_s\rightarrow \pi^+(\phi
\rightarrow K^+K^-)\sim 2\%)$.}
\end{quote}
\vskip 0.1in
\begin{tabular}{|| l | c | c |c ||}
\hline
Mode  & Br & yield$/10^{10}$ $B$'s & yield$/100 \; pb^{-1}$ \\ \hline \hline
$B_c\rightarrow \pi^+\psi \; $
& $4.0\times 10^{-3}$ & $2.8\times 10^3 $ $^{(a)}$  & $276$ $^{(a)}$ \\ \hline
$B_c\rightarrow D^\star_s \psi \; $ & $5.0\times 10^{-2}$
 & $7.0\times 10^2$ $^{(b)}$
 & few $^{(b)}$ \\ \hline
$B_c\rightarrow \psi \ell \nu \;$
& $10\%$ & $7.0\times 10^4$ $^{(a)}$ & \\ \hline
\end{tabular}
 \end{center}

Perhaps the most interesting new mesonic system will be the
$B_c = (\overline{b}c)$.  This is remarkable because we can say with
certainty that non--relativistic potential models apply, and the spectrum
is completely determined by those methods. Indeed, this is the true
Hydrogen atom of QCD.
Eichten and Quigg$^{13}$ have estimated the spectrum and widths of the $B_c$
system.  They use the Buchm\"{u}ller--Tye potential as fit
to the $\psi$ and $\Upsilon$ systems (and use other
potentials, e.g., the Cornell and Richardson potentials,
for error estimation), finding:
\beq
M_{Bc} = 6258\pm 20\;\;MeV\qquad
M_{B_c*}-M_{Bc}=73\pm \; MeV
\eeq
The prospects for observation of $B_c$ hinge upon the
production cross-section. There is reasonable agreement amongst
several groups$^{14}$  that
the ratio $\sigma(B_c)/\sigma(\overline{b}b)\sim 10^{-3}$ Thus, for $|\eta|\leq
1$ and $p_t>10$ GeV/c we have
$\sigma(B_c) \sim 10^{-2}$ $\mu b$, and a yield of
$\sim 10^7$ $B_c$'s for $10^{10}$ $B$'s.
Some of the principal detectable decay modes are listed in Table
III.$^{12}$

Note that the decay mode
$B_c\rightarrow \psi \ell \nu $ is the $B_c$  analogue of the
$B_u\rightarrow D^\star \ell \nu $ decay for which the
Isgur--Wise function at $w=1$ sets the normalization. Here the
process is completely determined, and $w=1$ involves only the
overlap of the known  $\psi$ and $B_c$ wave--functions.  Thus, this
is an interesting toy laboratory for the heavy quark symmetry
methods where everything is perturbative.
We should also mention that processes
containing $CP$--violation, like $B_c \rightarrow D_s\phi$, involve
both a direct short--distance penguin and interference terms with
short--distance contributions to
the imaginary parts.  Here the factorization approximation is
exact, and the short--distance imaginary parts are also in principle
computable. Thus, $CP$--violation in the $B_c$ system may ultimately
prove to be a fundamental issue in the $B$--physics program.
The $B_c$ is a remarkable system in which much of
the QCD dynamics is solvable by perturbative methods.  It will thus
provide a powerful laboratory for theorists and experimentalists,
and possibly a probative system  for new physics in the future.

\vglue 0.3cm
\leftline{\twelveit 2.7. Heavy Baryons}
\vglue 1pt

The spectroscopy and interactions of baryons consisting of two heavy
quarks and one light quark simplify heavy quark
mass limit, $m_Q\rightarrow \infty$.  The heavy quarks are
bound into a diquark whose radius $r_{QQ}$ is much
smaller than the typical length scale $1/\Lambda$ of QCD.
In the limit $r_{QQ}\lta 1/\Lambda$ the heavy diquark has
interactions with the light quark and other light degrees of freedom
which are identical to those of a heavy antiquark. Hence as far as
these light degrees of freedom are concerned, the diquark is nothing
more than the pointlike, static, color antitriplet source of the
confining color field in which they are bound, i.e., these $QQq$
baryons are in a sense ``dual'' to heavy mesons $\overline{Q}q$.

The spectrum of $QQq$
baryons is thus related to the spectrum of mesons
containing a single heavy antiquark.  The groundstate is
essentially a $(1,\half)$ or $(0,\half)$ heavy spin
multiplet. The form factors describing the semileptonic decays of these
objects may be directly related to the
Isgur-Wise function, which arises in the semileptonic decay of heavy
mesons.  The
production rates for baryons of the form $ccq$, $bbq$ and $bcq$
have been estimated in the approximation that the $QQ$ diquark is
formed first by perturbative QCD interactions, and then this
system fragments to form the baryon like a heavy meson.$^{15}$
(In the $cc$ system the heavy
diquarks are not particularly small relative to $1/\Lambda$, so there may
be sizeable corrections to these results).
Essentially the fragmentation of a heavy quark $Q$ into a $QQq$ (or $QQ'q$)
baryon factorizes into short-distance and long-distance contributions.  The
heavy quark first fragmentation into a heavy diquark  may be trivially
related to the fragmentation of $Q$ into quarkonium $Q\overline{Q}$.
This initial short distance fragmentation process  is
analogous to fragmentation into charmonium, $c\to\psi c$,
which has been analyzed recently by Braaten, et al., and others$^{14,15}$
 The subsequent fragmentation of the diquark $QQ$ to a baryon is
identical to the fragmentation of a $\overline{Q}$ to a meson
$\overline{Q}q$.$^{15}$
Experimental data on production of
heavy mesons can be used here.

\vskip 0.1in
\begin{center}
\begin{quote} {\tenrm Table IV. Hadronically  produced double heavy baryons
for Tevatron ($ 3\times 10^{9}$ $B_{u,d}$'s)
and Main Injector  ($3\times 10^{10}$ $B_{u,d}$'s).}
  \end{quote}
\vskip 0.1in
\begin{tabular}{|| l | c | c | c | c ||}
\hline
Mode  & Tevatron  & Main Injector \\ \hline \hline
$\Sigma_{cc}, \Sigma_{cc}^*$ & $6\times 10^4$  & $6\times 10^5$  \\ \hline
$\Lambda_{bc}$ & $ 6\times 10^{4}$ & $6\times 10^{5}$  \\ \hline
$\Sigma_{bc},\Sigma_{bc}^*$ & $\sim 10^5$ & $\sim 10^6$ \\ \hline
$\Sigma_{bb},\Sigma_{bb}^*$ & $\sim 10^3$ & $\sim 10^4$ \\ \hline
$\Lambda_{bc}$  & $ 6\times 10^{2}$ & $6\times 10^{3}$  \\ \hline
$\Sigma_{bc},\Sigma_{bc}^*$ & $6\times 10^2$  & $6\times 10^3$  \\ \hline
\end{tabular}
\end{center}
\vskip 0.1in

The
probability for $c\to\Sigma_{cc}, \Sigma_{cc}^*$ is estimated to be
$\sim2\times10^{-5}$, for $b\to\Lambda_{bc}$ to be
$\sim2\times10^{-5}$, and for $b\to\Sigma_{bc},\Sigma_{bc}^*$ to be
$\sim 3\times10^{-5}$.  The probabilities for
$b\to\Sigma_{bb},\Sigma_{bb}^*$, $c\to\Lambda_{bc}$ and
$c\to\Sigma_{bc},\Sigma_{bc}^*$ are down by roughly $(m_c/m_b)^3$, or
two orders of magnitude.

Detection of these objects is probably very difficult at best.
Consider the $\Sigma_{bb}$ decay chain:
\bea
\Sigma_{bb}& \rightarrow & D^* + X + (\Sigma_{bc}
\nonumber \\
&\qquad & \qquad  \rightarrow D^* + X + (\Lambda_{b}
\nonumber \\
&\qquad&  \qquad\qquad  \rightarrow  D^* + X + (\Lambda_{c}
\nonumber \\
&\qquad&  \qquad\qquad\qquad   \rightarrow  K^* + X +\Lambda
\eea
Each vertex above must be reconstructed, in spite of a high probability of
missing neutrals, including
the drift of $ D^*\rightarrow D$'s away to branch vertices.  A rough estimate
is that a handful of such
decay chains might be available in a $10^{10}$ program admitting
reconstruction of the parent doubly--heavy baryon.  However,
there will come insights as to how to do this well
as experience is gained.

\section{Rare Processes }

In this section we will briefly discuss some of
the interesting ``rare'' processes that are the
far--reaching goals of the  initiatives of this decade.  Much
greater detail is afforded these topics
in other talks in this conference, so we will focus only
on issues that involve  some of the aforementioned
ideas.  Clearly the ultimate structure of
$CP$--violation is of great interest, but the first observation
of $CP$--violation in the $B$-system will be an achievement
of enormous importance.  We will comment as to how
this observation may be feasible in the hadronic collider
mode by making use
of daughter pion flavor tagging, in comparison to the conventional
strategy.  Indeed, many of the tools necessary to see
the $CP$--asymmetry in $B\rightarrow \psi K_S$ are now in place
at CDF,
and this exciting observation may be only a few years away!

We describe the important observation of $B_s\overline{B}_s$
mixing.  This process will be quite a bit
more difficult to
observe than $CP$--violation.
This is likely, given that the large top mass implies a large $x_s$,
and mandates very high statistics for flavor tagged, and kinematically
tagged $B_s$ semileptonic decays.  It may be a leap of faith to
extrapolate to this process, given that there is limited experience
with semileptonic decays of any $B$--system to date. In conjunction
with flavor tagging, our experience here is $O(\epsilon^2)$ at
present.
We will also discuss the rare leptonic modes.  Here we have
made extensive use of a presentation by S. Willenbrock and G. Valencia
from our in--house workshop. Thus, the last subsection is really their
effort, more than mine.

\vglue 0.3cm
\leftline{\twelveit 3.1. CP violation  }
\vglue 1pt

There are well--known modes for the observation of
$CP$--violation, such as $B^0\rightarrow \psi \; K_S$, etc.,
and $B_s\rightarrow D_s^\pm\; K^\mp$, and self--tagging modes.$^{16}$  To
observe $CP$--violation
we must tag the flavor of
the initial state, which taxes the available statistics.
$CP$--violation with self--tagging modes is experimentally attractive,
but there exists no guarantee that observable $CP$--effects will be present
in these modes.$^{16}$  Since the volume of the Snowmass Proceedings is
consumed with the intimate details of $CP$--violation
in the $B$--system, we will simply focus on how one might use
the conventional or daughter--meson tagging methods to observe
the straightforward $B^0\rightarrow \psi \; K_S$ $CP$--asymmetry.

The decay mode $(B^0,\;\overline{B}^0) \rightarrow \psi \; K_S$ involves
$CP$--violation.
Thus the partial widths for $B^0$ and $\overline{B}^0$ to decay
into the $\psi \; K_S$ final state differ, and the time integrated asymmetry
is defined as:
\beq
 a = \frac{\Gamma(\overline{B}\rightarrow \psi K_S)
- \Gamma({B}\rightarrow \psi K_S)}{\Gamma(\overline{B}\rightarrow \psi K_S)
+ \Gamma({B}\rightarrow \psi K_S)}
= \frac{x_d}{1+x_d^2}\sin(2\beta) \sim 0.1 \; - 0.5
\eeq
Note that the branching ratio for $B^0\rightarrow K_S+( \psi
\rightarrow \mu^+\mu^-) $ is $\sim 2\times 10^{-5}$
(including the $7\%$ dimuon mode of $\psi$).

To observe $a$
one must flavor--tag the neutral $B$--meson at production $t=0$
to determine if it is a particle or
anti--particle. Since $b$--quarks are produced
in pairs, this is conventionally
achieved by observing the  semileptonic decay mode of
the other $B$ in the event.  For example, if the
other meson is a $B^-$ ($\overline{B}^+$) it can decay
semileptonically to a charge $-$ ($+$) lepton, with
a $Br(B\rightarrow \ell\nu D)\sim 10\%$.  This does not require full
reconstruction of the semileptonic decay, so for
$CP$--violation one
is effectively measuring
$\Gamma(\ell^+\psi K_S) - \Gamma(\ell^-\psi K_S)$ (Note that this does not
require a new single lepton trigger since one can
trigger on the  $\psi$ dimuons).
Including geometric efficiencies this
conventional tagging efficiency is expected to be
of order $\epsilon_1\sim 10^{-2}$.

 Gronau, Nippe and
Rosner$^3$ have pointed out that resonance daughter pions
(as well as rapidity correlations associated
with the jet fragmentation) are possible flavor--tags.
A stunning implication of the daughter mesons from parent
resonances is that all $CP$--violating processes in hadron
machines are expected to be self--tagging!
We should recognize that at low--$p_T$ the
$b$--production mechanism is
somewhat more akin to threshold production and the resonance
mechanism may be favored.

\vskip 0.1in
\begin{center}
\begin{quote}
{\tenrm Table V. Statistical significance $\sigma_i$ for tagging efficiencies
$\epsilon_1,\epsilon_2$ and asymmetries $a$, for various integrated
luminosities.
We show the $100\; pb^{-1}$, i.e.,  prospects for run I(b) at Fermilab
($10^{10}$ $B$'s corresponds to $\int{\cal{L}}dt=10^3$ $pb^{-1}$).}
\end{quote}
\vskip 0.1in
\begin{tabular}{|| l | c | c | c ||}
\hline
$a$  & $\epsilon_2 -\epsilon_1$  & $\int{\cal{L}}dt$ & $\sigma_2 - \sigma_1$ \\
\hline \hline
$0.5 $ & $0.1 - 0.01 $  & $100\; pb^{-1}$ & $ 2.1 - 0.7 $ \\ \hline
$0.1 $ & $0.1 - 0.01 $  & $100\; pb^{-1}$ & $ 0.4 - 0.13 $ \\ \hline
$0.5 $ & $0.1 - 0.01$  & $10^3\; pb^{-1}$ & $ 6.7 - 2.1 $ \\ \hline
$0.1 $ & $0.1 - 0.01 $  & $10^3\; pb^{-1}$ & $ 1.3 - 0.4$ \\ \hline
$0.5 $ & $0.1 - 0.01$  & $10^4\; pb^{-1}$ & $ 21.2 - 6.6 $ \\ \hline
$0.1 $ & $0.1 - 0.01 $  & $10^4\; pb^{-1}$ & $ 4.1 - 1.3 $ \\ \hline
\end{tabular}
\end{center}
\vskip 0.1in

\noindent
At higher $p_T$ the $b$--jet is forming
and there would be more pions expected (a source of dilution),
and perhaps the rapidity correlation idea is favored. This is not
to advocate any theory of production, but rather to emphasize that the
optimization may involve tuning of $p_T$, etc. For example, we
may prefer operating at low $p_T$'s below the present cuts.
While with optimization cuts it is possible that significant
improvements in the tagging efficiency may occur,
the charm photoproduction experiments suggest that a tagging efficiency of
$\epsilon_2\sim 10\%$
from daughter pions is possible. The flavor of a neutral
$B^0\sim \overline{b}d$   ($\overline{B}^0\sim {b}\overline{d}$)
is tagged by the presence of a $\pi^+$ ($\pi^-$) daughter,
and the $CP$--asymmetry we measure in practice is
effectively $\propto \Gamma(\pi^+\psi K_S) - \Gamma(\pi^-\psi K_S)$.

The overall efficiency for observing $B\rightarrow K_s(\psi\rightarrow \mu\mu)$
involves the
physics branching ratio $\sim 2 \times 10^{-5}$ times the detection
efficiency (including geometric efficiencies).  The latter is $\sim 3\%$ at
the CDF SVX at present, and we assume it in Table IV. Thus, the overall
efficiency for $B\rightarrow K_s(\psi\rightarrow \mu\mu)$ is
$\sim 6\times 10^{-7}$, and, for $100$ $pb^{-1}$, we expect $3\times 10^{8}$
usable neutral $B$'s, therefore $\sim 180$ $\psi K_s$ events.  Larger
$\eta$ coverage, and other detector gains might boost this $\sim 5\times $.

The prospects for observing the CP-asymmetry at a statistical deviation
$\sigma$ are indicated in Table V.
Significant limits on $CP$--violation in the $B$
system will begin to be placed
by end of run I.
In the best case, $a=0.5$
we can begin to see a signal with the conventional
semileptonic tagging efficiency, $\epsilon = 0.01$, for
$10^{10}$ produced $B$'s, or with the
daughter pion tagging $\epsilon_2 = 0.1$ and the larger asymmetry a
discovery is likely. Evidently a discovery is
assured for $10^{11}$ $B$'s with daughter pion tagging.

\vglue 0.3cm
\leftline{\twelveit 3.2. $B_s\overline{B}_s$  Mixing}
\vglue 1pt

We have for the mixing parameter:
\bea
x_s & = & \frac{G_F^2m_{B_s}\tau_{B_s}}{6\pi^2}B_sf^2_{B_s}\eta_B
|V^*_{ts}V_{tb}|^2m_t^2 F(m_t/M_W)
\nonumber \\
&\approx &\Delta M_{B\overline{B}}/\Gamma
\sim (14\pm 6)(f_{Bs}/200\;MeV)^2
\eea
where $F(z)$ is an Inami-Lim function. An expression for
$x_d$ is gotten by replacing $s$ by $d$ everywhere.
Note that:
\beq \frac{x_s}{x_d} = \left|\frac{V_{ts}}{V_{td}}\right|^2 (1 + \delta )
\qquad
\delta = \left(\frac{m_{B_s}f^2_{B_s}}{m_{B_d} f^2_{B_d}} -1\right)
\sim 0.2\;
\eeq
$x_s$ is very sensitive to
$m_{top}$ and we find:
\bea
x_s \sim 8.0\; \leftrightarrow 18.0, \qquad& & m_t=140\; GeV;\;
\sqrt{B}f_B=200\; MeV \\ \nonumber
x_s \sim 17.0\; \leftrightarrow 40.0, \qquad & & m_t=200\; GeV;\;
\sqrt{B}f_B=220\; MeV \\ \nonumber
\eea
and we must prepare ourselves for the possibility
of large $x_s$, $8\lta x_s \lta 40$.
For large values the system oscillates many times per decay length
($x = \half$(radians)/(e-attenuation), thus $x=10$
corresponds to  $20$ radians
per decay length).  This requires observing the time evolution of
the system, which implies that fully reconstructed (energy and flavor),
tagged $B_s$ decays are necessary.
In contrast, $x_d = 0.66$ and is readily observed
in time--integrated measurements.   These requirements make the observation
of $B_s\overline{B}_s$  mixing more challenging than the observation
of $CP$--violation!  However, it should be emphasized that this important
phenomenon is likely to be the exclusive province of hadron collider
experiments because of the large statistical requirements.

The key to observing oscillations is achieving the smallest
proper time resolution, $\sigma_t$  (for a good schematic discussion
of this see Mike Gold in ref.(1); we also thank John Skarha for
discussions on this topic).  $\sigma_t$ is composed
of the
beam-spot resolution $\delta L_{xy}/L_{xy}$ where $L_{xy}$ is the
transverse path length (this is the
dominant contribution), together with the momentum resolution
$\delta p_T/p_T$ as:
\beq
\sigma_t = \left( (\delta L_{xy}/L_{xy})^2 + (\delta p_T/p_T)^2\right)^\half
\eeq
With $\delta_{xy}\sim 40\;\mu m$, $L_{xy}\sim 600\; \mu m$,
we find $\sigma_t\sim 0.07$ characteristic of CDF-SVX.

The conventional triggers would  use a produced
$B_s\rightarrow {{l}}\nu (D_s\rightarrow \phi X)$
or  $B_s\rightarrow \pi^+\pi^-\pi^+(D_s\rightarrow \phi X)$
and the opposite $B\rightarrow {{l}}\nu X$ for
flavor tagging. By fully reconstructing the $B_s$
decay (requiring exclusively  charged particles in $X$)
and partially reconstructing the tagging
decay, it has been estimated
that one can reconstruct the oscillation in $\tau$
with $\sim 4000$ events.$^1$  With the estimated
efficiencies this requires about $3\times 10^{10}$ to $10^{11}$
produced $B$'s.
This appears to be a significant challenge!

Can we tag the $B_s$ flavor and kinematics by using the
daughter $K$ mesons associated with it's resonance production?
For example, we expect the D-wave $B(2^-)$ and $B(3^-)$ to be above
threshold for decay to $K^+ + B_s$ or $K^- + \overline{B}_s$.
These resonances are estimated to be broad ($250$ to $400$ MeV),
but with a decay fraction to $B_s$ and $B^*_s$ of about $30\%$.
Thus, with the favorable production and branching fractions we may have a
flavor tag for $B_s$, but a kinematic tag seems less likely.
The charm system process $D_s^{**}\rightarrow D^*K$ has been
demonstrated,$^5$ which is the opposite to $D_s^{***}\rightarrow D_sK$,
The higher resonances have not yet been seen.

\vglue 0.3cm
\leftline{\twelveit 3.3. Other Rare Modes}
\vglue 1pt

Length considerations preclude our giving any comprehensive
discussion of the additional interesting rare modes in $B$-physics.
We will, however, briefly mention a few of the leptonic modes.
Rare $B$ decays encompass such processes as:
\bea
(I) \qquad B_{d,s}& \rightarrow  & (e\overline{e},\;
\mu\overline{\mu},\;\tau\overline{\tau})
\nonumber \\
(II) \qquad B_{d,s}& \rightarrow  & (e\overline{\mu},\;
\mu\overline{\tau},\;e\overline{\tau})
\eea
and additional hadrons in the final state may be included.
We should remark that the $\tau$ containing final states
are unique to $B$, never available in $K$ decays, and at best
phase space suppressed for $D$'s.

Such processes as $(I)$ have low standard model rates and
are probes of $V_{td}$, $V_{ts}$, and $m_t$.  Thus, they are
good probes of the standard model if they are seen at
the expected rates.  Moreover, they are excellent probes of new
physics, such as charged Higgs and flavor changing neutral
Higgs couplings, which are generally $\propto $ mass. The
conventional SM estimates are as follows:

\vskip 0.1in
\begin{center}
\begin{quote}
\begin{center}
 {\tenrm Table VI. Rare leptonic mode branching ratios.}
\end{center}
\end{quote}
\vskip 0.1in
\begin{tabular}{|| l | c | c | c ||}
\hline
  & $\tau \overline{\tau} $ &
$ \mu \overline{\mu} $ & $e \overline{e} $ \\ \hline \hline
Br$(B_s \rightarrow \;)$ & $10^{-7}$ & $10^{-9}$ & $10^{-14}$ \\ \hline
Br$(B_d \rightarrow \;)$ & $5.0\times 10^{-9}$ & $5.0\times 10^{-11}$ &
$5.0\times10^{-16}$ \\ \hline
\end{tabular}
\end{center}
\vskip 0.1in

A crude estimate of the
background due to Valencia and
Willenbrock is as follows.  UA-1 has measured the continuum
$\mu$--pair background cross--section near the $B$--mass, $M_{\mu^+\mu^-}^2 =
M_B^2$
with momentum resolution $\delta p \sim 100 $ MeV to be
$\sigma({\mu^+\mu^-}) \sim 10^{-5}\sigma_B$, where $\sigma_B$ is the
hadronic $B$ cross--section.  This can presumably be reduced to
$\sigma({\mu^+\mu^-}) \sim 10^{-6}\sigma_B$ with improved momentum
resolution from silicon vertex detectors. The probability that
two stray muons make a vertex is geometrically $\sim 10^{-2}$ and
the probability that this yields a momentum vector pointing toward
the primary vertex is $\sim 10^{-2}$.  Thus we have an overall
background approaching $\sim 10^{-10}\sigma_B$ and a 3--$\sigma$
$B_s$-peak is therefore possible.  With a yield of
$10^{11}$ $B$'s we expect therefore $\sim 30$ events from $B_s
\rightarrow \mu\overline{\mu}$.  Since the signature is a clean
displaced muon pair event with mass reconstruction, it is likely
that this can be searched over a rapidity range of $|\eta|\lta 3$,
and a $p_t$ threshold of $O(5)$ GeV/c.

Valencia and Willenbrock (VW) have given a nice characterization
of the lepton--number violating processes (class II, above) which we describe
here.  First, note that $(\tau\;e)$ and  $(\tau\;\mu)$
are unique to the $B$-system (not available in rare $K$ decays).
Since such processes can be generated in principle by Higgs-scalar
exchange, which is a coupling constant $\propto$ mass, it is possible
that the $B$ system becomes sensitive to these processes at a level
that is readily experimentally accessible, and complimentary to rare $K$
decay searches, such as at KTEV.

VW begin by postulating general four--fermion
interactions describing such processes as $B\rightarrow e\mu$ and
$K\rightarrow e\mu$ as:
\beq
c_B(\overline{s}\Gamma d\;\overline{\mu}\Gamma e) + c_K(\overline{b}\Gamma
s\;\overline{\mu}\Gamma e)
\eeq
with arbitray Dirac structures $\Gamma$.  VW then consider the
effects of different $\Gamma$'s and $c_X$'s on the ratio
of branching ratios $R_1= Br(B\rightarrow \mu e)/Br(K_L\rightarrow \mu e)$
and $R_2= Br(B\rightarrow \mu e + h)/Br(K_L\rightarrow \mu e+h)$
(where $h$ is an extra hadron system, e.g., pions) as follows:
\bea
R_1 &\approx& \frac{c_B^2f_B^2}{c_K^2f_K^2}\left(
\frac{m_B\tau_B}{m_K\tau_K}\right) \approx 10^{-4}\frac{c_B^2}{c_K^2}
\qquad \Gamma =(\gamma_\mu, \gamma_\mu\gamma^5)
 \nonumber \\
R_1 &\approx &\frac{c_B^2f_B^2}{c_K^2f_K^2}\left(
\frac{m^3_B\tau_B}{m^3_K\tau_K}\right) \approx 10^{-2}\frac{c_B^2}{c_K^2}
\qquad \Gamma =(1, \gamma^5)
 \nonumber \\
R_2 &\approx &\frac{c_B^2f_B^2}{c_K^2f_K^2}\left(
\frac{m^5_B\tau_B}{m^5_K\tau_K}\right) \approx \frac{c_B^2}{c_K^2}
\eea
Thus, to proceed we need input as to the magnitude of the
ratio $c_B/c_K$.  VW distinguish three cases:
(i) (Current-like) $c_B/c_K \sim 1$ (ii) (Higgs-like)
$c_B/c_K \sim m_B/m_K \sim 10^1$  (iii) (Box--like) $c_B/c_K \sim V_{tb}V_{ts}/
V_{td}V_{ts} \sim 10^2$.  The latter ``box-like'' result assumes that
the process is induced via a top quark containing box diagram.
Thus, the following table arises:

\newpage
\begin{center}
\begin{quote}
 {\tenrm Table VII. Valencia and Willenbrock's characterization of
lepton--number violating modes of $B$ and $K$.}
\end{quote}
\vskip 0.1in
\begin{tabular}{|| l | c | c | c ||}
\hline
$Br(B\rightarrow X)/Br(K\rightarrow X)$ &
 Current-like & Higgs-like &  Box-like
 \\ \hline \hline
$\Gamma =(\gamma_\mu, \gamma_\mu\gamma^5)$; $X=e\mu$ & $\sim 10^{-4}$ & $\sim
10^{-2}$ & $\sim 1$ \\ \hline
$\Gamma =(1, \gamma^5)$; $X=e\mu$ & $10^{-2}$ & $1$ & $10^2$ \\ \hline
Any $\Gamma$; $X=e\mu +h$ & $1$ & $10^2$ & $10^4$ \\ \hline
\end{tabular}
 \end{center}
\vskip 0.1in

Thus, in the ``box--like'' and ``Higgs--like'' limits the
$B$ system maybe a better probe than the $K$ system for
new physics.
The VW characterization is general, and covers all possible models.
It illustrates the possibility that $B$ decays are sensitive to new
physics in a manner complimentary to rare $K$'s.

\section{Summary}

A program of producing $>10^{10}$ detectable
$B$'s is conservatively achievable within this
decade.  This offers an excellent
conventional physics program of $\sim 10^9$ $B\rightarrow D^*\ell \nu$
decays and $\sim 10^5$ $B\rightarrow \rho\ell \nu$ decays, allowing
a determination of $V_{cb}\pm 3\%$ and $V_{ub}\pm 20\%$. This also probes
the quantities such as $\sqrt{B}f_B$ and $f_{B_s}$ with high statistics.

The resonances and the prospects for flavor and kinematic tagging
will emerge within the next few years. New states such as
$B_c$ will be surveyed, and the list of $B_s$ and $B_c$ decay modes
will grow.
$CP$--violation with conventional or bachelor pion tagging may be
first observed in the $\psi K_S$ asymmetry within such a $10^{10}$ program.
$B_s\overline{B}_s$ mixing looks difficult, though $x_s\lta 20$ may
be probed.  Rare and radiative decays will be subject to their
first probative examination.

In conclusion, we have seen that $B$--physics based
in a hadron collider offers a rich and diverse, unique and powerful
scientific program.  It can peacefully coexist with a high--$p_T$
program and dominate the post--High--$p_T$ era at such facilities as
Fermilab.  Indeed, the prospects for observation of $CP$--violation
in the $p\overline{p}$ collider environment are great. There are
in fact advantages of the  $p\overline{p}$ mode over $p{p}$
in the observation of $CP$--violation.  A dedicated $B$--physics
program at Fermilab is important to the evolution of the world--wide effort
and a healthy base program for at least the next ten years and probably
beyond.

\newpage
\section{Acknowledgements}

This work was performed at the Fermi National Accelerator
Laboratory, which is operated by Universities Research Association,
Inc., under contract DE-AC02-76CH03000 with the U.S. Department of
Energy. Much of it
has been influenced strongly by an in--house $B$--physics working group at
Fermilab which we convened in March `93, and which met approximately weekly
throughout the spring.   I wish to thank CDF and D0 experimentalists, in
particular Fritz DeJongh, Dan Green,
Jim Mueller, John Skarha, Paris Sphicas,
and especially Tom LeCompte,
who were active contributors in this group. Also, I wish to thank my
theoretical
colleagues, I. Dunietz, E. Eichten, R. Ellis, A. Kronfeld, C. Quigg,
J. Rosner, J. Soares,
G. Valencia and S. Willenbrock  for numerous contributions
and discussions.
I also thank A. Yagil and S. Stone for several useful discussions
and C. Quigg for proof--reading an earlier version.
This view of a $B$--physics program was subsequently presented to the Fermilab
PAC.  My present talk is essentially that sketch, embellished with a lot of
personal biases.

\end{document}